# Inverse Modeling of Complex Networks Using Embedded Complex Logistic Maps

Sandy Shaw*
*Fractal Genomics, LLC*
*San Francisco, CA 94118*
*sandy@fractalgenomics.com*

**Abstract**

An inverse modeling technique is introduced that combines elements of coupled logistic map models and wavelet analysis for the purpose of analyzing partial synchronization states in high-dimensional systems. Using Embedded Complex Logistic Maps (ECLM), time series data derived from individual system components is directly mapped to a wavelet-like space generated from iterations of specific complex logistic maps. These maps are selected from the complex plane according to "best fit" scoring criteria with the data. The embedding topology within and near the familiar Mandelbrot Set provides metrics which are used to aid in clustering similarities (synchronization) between these selected point models. The dynamics within the individual models and the correlation between (synchronized) models is analyzed to reconstruct a unified picture of the underlying dynamics of local system components within the global system network topology. In this paper, ECLM is used to extract system parameters, network graphs, and wavelet-like analytics from two real-world systems, a gene expression network and a financial market network. Preliminary results appear to validate the assumptions and methods used in ECLM through agreement with current theory and recent findings. Some potentially new findings regarding synchronization, scale-free networks, on-off intermittency, and energy dissipation within the examples studied will also be discussed.

## I. Introduction

Coupled logistic maps [1,2] have been used to successfully model the dynamics of a variety of collective phenomena present in high-dimensional systems such as partial synchronization, power-law connectivity, and on-off intermittency [13-15, 17]. The ability to model high-dimensional systems with such low dimensional models appears to be aided by the properties of synchronization and long-range order, which help to effectively lower the dimensionality of the system [3, 7]. Despite these successes, and the effective lessening of the "curse of dimensionality" [6] brought about by these effects, reconstruction of the dynamics of complex, high-dimensional systems is a difficult task requiring far more data than can generally be acquired by experiment [5].

Wavelet analysis has been used to model aspects of high-dimensional systems directly from time series data [4]. A kind of localized Fourier Analysis, wavelet analysis is particularly effective in modeling transient or "bursty" behavior of systems [8]. Such behavior is present in a number of complex networks and systems, such as Internet traffic, financial market data, heartbeat time series, EEG data, etc. [8-10].

The method of Embedded Complex Logistic Maps (ECLM) is a hybrid analysis method that blends some of the concepts behind these two approaches to deal with the inverse modeling problem. ECLM is based on point models within the complex plane that are used to form complex logistic maps (Julia sets [12]). Rather than couple these models together, they are placed at particular points in the complex plane based on how their iterations, or orbits [12], best model individual system components. The models are then compared in their "natural" topology, within and near the Mandelbrot Set [11]. This comparison is aided by metrics on the surface that are used to cluster similarities between the models [64, 67, 74]. By creating individual models and performing global comparisons in this way, ECLM is able to give a picture of local element dynamics in a global framework.

Like wavelet analysis, ECLM is well suited to probe for transient behavior. In particular, ECLM is designed to probe for locally transient synchronization within coupled chaotic systems in the so-called "glassy" or partially ordered state [13, 14]. In this state the system exhibits local clusters of partial synchronization and on-off intermittency [15]. In states with on-off intermittency, episodes of nearly synchronous evolution are interrupted by elements in the system bursting away from the synchronous state. It is these transient synchronization states that are probed by ECLM as the system continuously forms synchronized pairs and clusters that split apart and reform in different ways. The motivation for probing these states is that the interplay of synchronized system elements can be used to gain insight into the local and global dynamics as well of the changing topology of the system being studied.

In this paper, ECLM will be applied to two real-word examples, a temporal gene expression study of Rat CNS development (Wen, et. al.) [16] and a group of stock market indices during the period April 2001-April 2002. For each of these examples we will analyze connectivity between system elements, on-off intermittency, synchronization cluster size distribution, and attempt to analyze energy dissipation in these dissipative (open) systems. This analysis will be compared with current theory and findings toward validation of ECLM as well as elucidating some potentially new findings in these real-world settings. Diagrammatic analysis and a type of ECLM "wavelet" analysis with predictive capabilities will be briefly outlined. Methodology for ECLM and computational efficiency will be detailed in a separate section (see Appendix "ECLM Methodology"). These preliminary results will be discussed as well as future directions of study.

## 2. ECLM Analysis of a Genetic Network

The input data for this ECLM trial consisted of gene expression data from 112 genes collected at nine different time points over 25 days, stretching from the embryonic stage through birth and post-birth, during rat central nervous system (cervical spinal cord and hippocampal) development [16]. The ECLM methodology for this experiment is described in the Appendix. In summary, each of the 9 time points (including 100 corresponding snapshots) were fitted to an ECLM point model with $|\mathbf{C}| > .95$, where **C** is the Pearson correlation coefficient (assuming such a model could be found). All the fitted point models were then compared to each other to see which ones had pairwise $|\mathbf{C}| > .95$ for the 9 (model) time points. This led to 69 genes with a least one link to another gene (pairwise $|\mathbf{C}| > .95$ with at least one other ECLM model). We call these models linking ECLM (LECLM). (See section "LECLM Analysis" for a view of plots of some representative LECLM).

### 2.1 Connectivity Distribution

Scale-free networks have been an area of much study in recent years [21]. In scale-free networks, **P(k)**, the likelihood that a randomly chosen node from the network has **k** direct interactions, decays as a power law, that is, free of a characteristic scale [20]. Some recent results have shown scale-free structure in cellular metabolic networks and protein interaction networks. Some very recent results have indicated scale-free structure in transcription regulatory networks (global RNA expression) [18] and in gene expression networks [19]. It has also been suggested that all gene expression follows Zipf's law [23]. From the perspective of dynamics, scale-free structure has recently been found to spontaneously appear in coupled logistic maps when linking was based on dynamical considerations [17] rather than the notion of *preferential attachment* [24] usually cited in the evolution of scale-free networks.

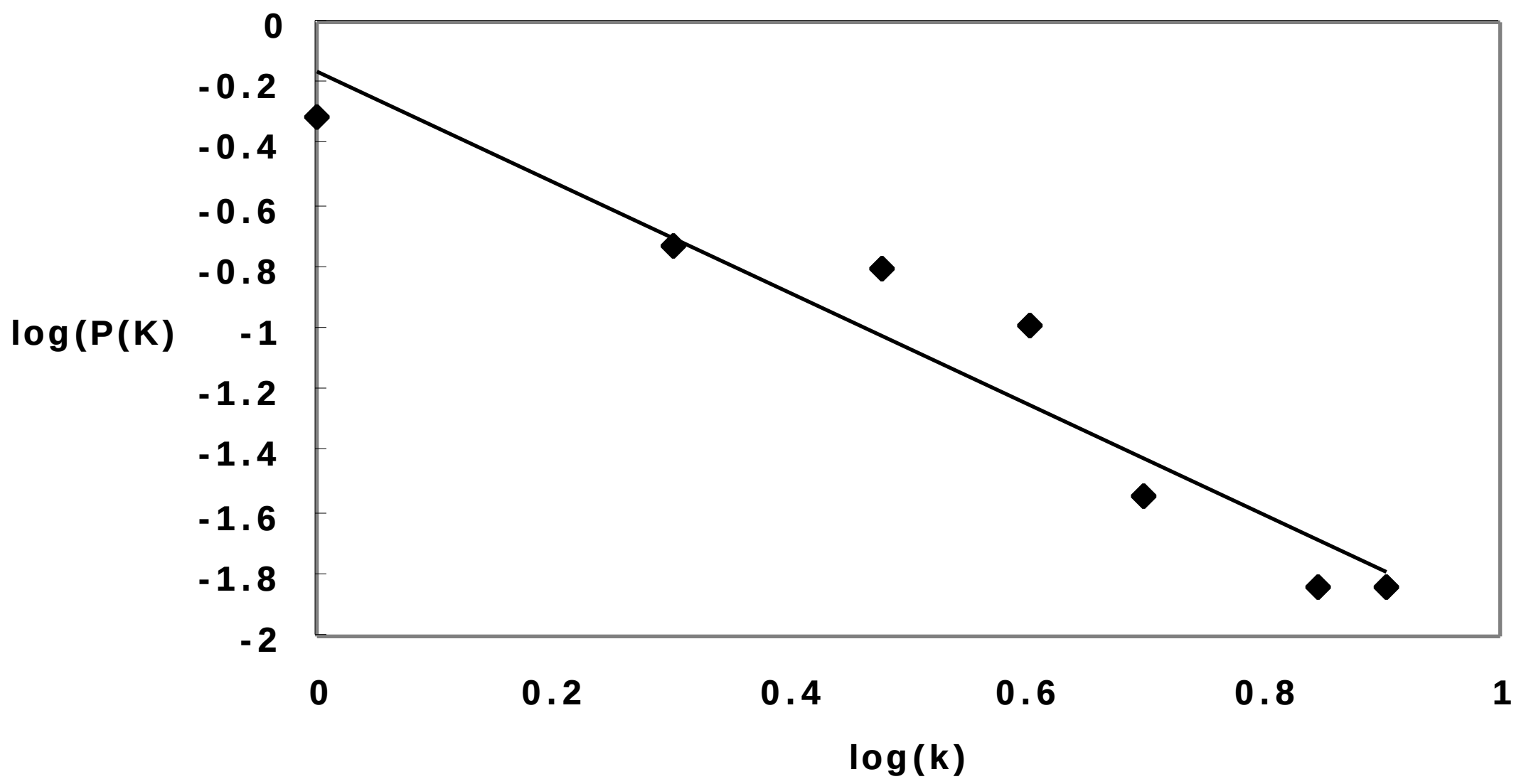

Fig. 1: Log-Log probability distribution for connectivity of linked ECLM models derived from genetic network data.

In Fig.1 you see the log-log plot of **P(k)** for the connectivity of LECLM models fitted to the temporal gene expression from the Wen study. Although the data sample is small, the apparent scale-free structure in the data (linear correlation >.95) lends support to the assumptions of the dynamical model in the coupled logistic map study referenced above and appears to agree with the recent findings in other biochemical networks. The scaling constant of ~1.75 falls in the range of scaling found in the metabolic and protein studies cited [25-28]. Besides lending support for the validity and usefulness of ECLM, this apparent scale-free connectivity is intriguing because it was produced from a dynamical perspective, using real data from a high-dimensional biological system, not just a model. The finding also has potential significance because the recent studies related to scale-free genetic networks were based on the organism S. cerevisiae, a yeast, and the Wen data is mammalian in origin.

### 2.2 On-Off Intermittency

When a system of chaotic elements is in the partial ordered (glassy) state, it has been shown to exhibit a chaotic phenomena known as on-off intermittency, where synchronization between elements in the system is interrupted by one or more of the elements bursting away from the synchronous state. This behavior has been shown to lead to synchronization times between elements with a power-law distribution over many decades [15]. This behavior has been found to exist in some real-world systems and coupled models [15, 29-31].

The LECLM models fitted to the 9 time points (see Appendix) were iterated to explore their asymptotic behavior. Of the 146 LECLM models found, only 10 were found to be bounded after 10000 iterations. For each of the unbounded LECLM, iterations were stopped when the size of the orbit went above a set cutoff value above which it was known to go infinity [11]. Fig. 2 is the distribution of the number of iterations required to reach the cutoff for all unbounded LECLM. One can interpret the number of iterations below the cutoff for each LECLM model as a rank measure of the (relative) duration for the transient synchronization corresponding to that particular model. With this interpretation, a distribution plot of the cutoff iteration values should show power-law scaling if on-off intermittency is actually being detected [15, 32].

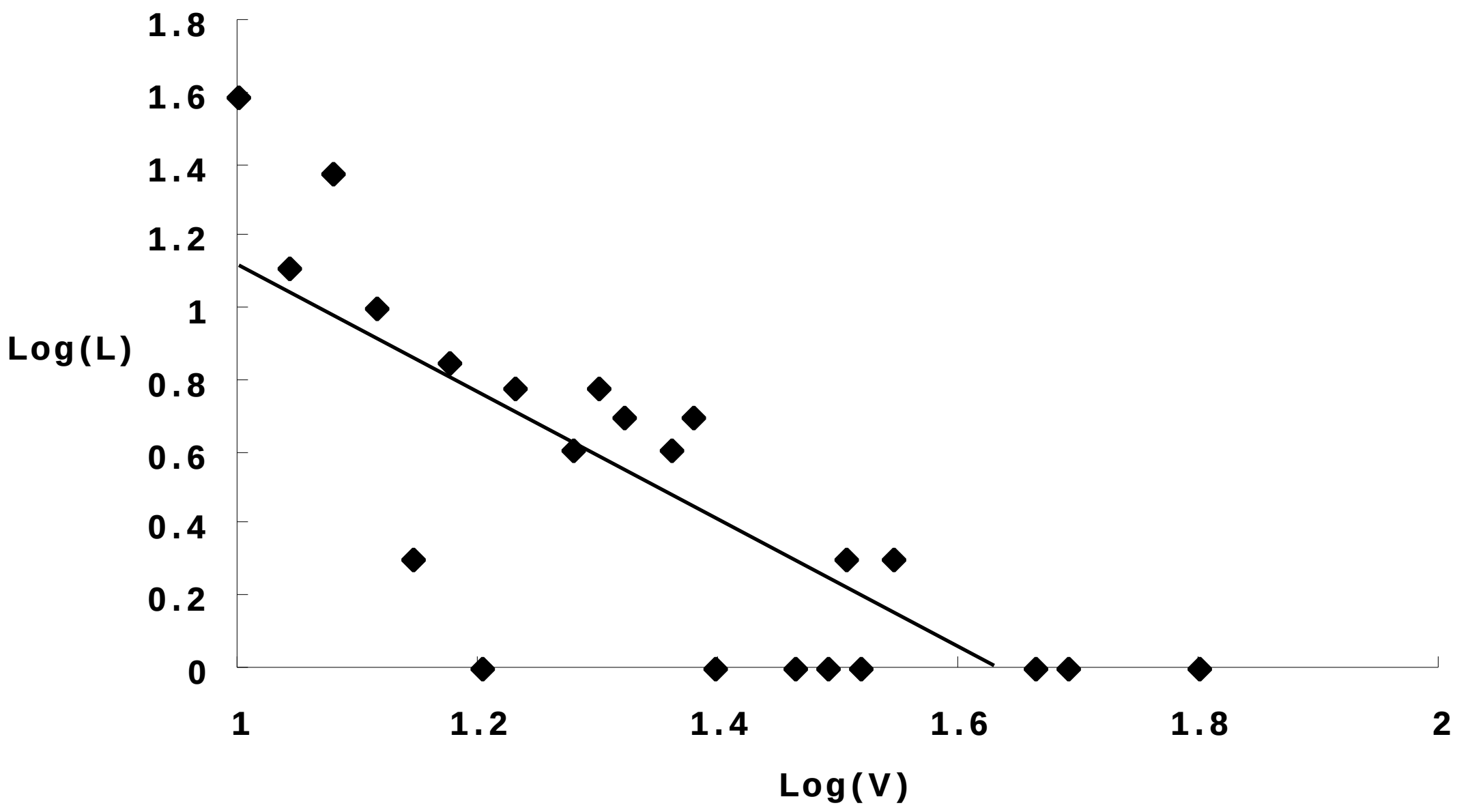

Fig. 2: Log-Log distribution where **V** = the unbounded LECLM cutoff iteration values and **L** = the number of unbounded LECLM with value **V**

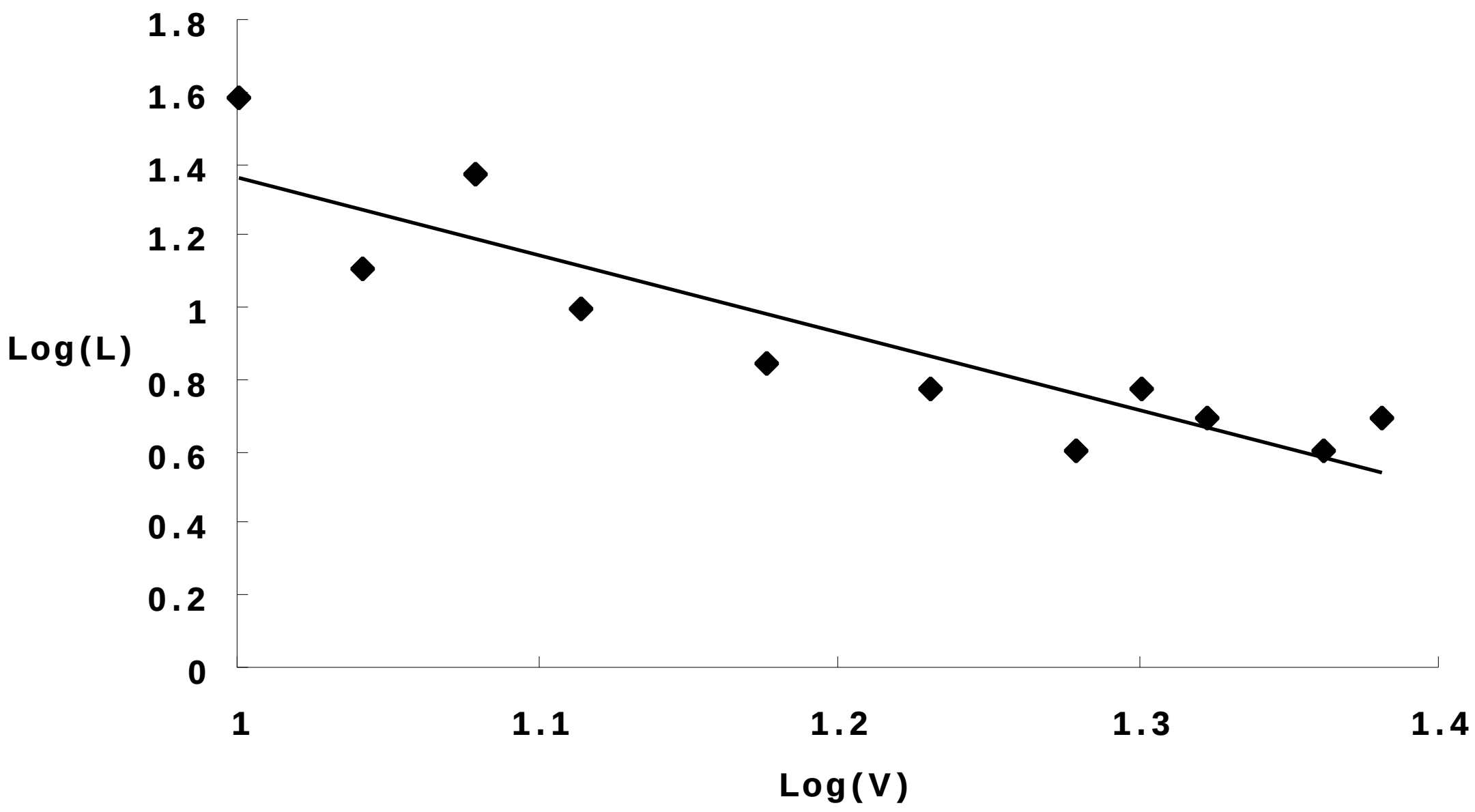

Fig. 3: Same as Fig. 2 with weakly populated states removed.

Fig. 2 is the log-log distribution plot of the LECLM cutoff iteration values **L** vs. the number of unbounded LECLM possessing that iteration cutoff value **V**. The scaling of the fitted line is ~1.8. There does appear to be some linearity in scaling for values of **V** that are possessed by more than 1 or 2 LECLM. Fig. 3 is the same plot with weakly populated states removed. The scaling of the fitted line is now ~2.2, with a linear correlation of ~.9. The sparseness of the data makes any assumptions about this result difficult but the appearance of some linearity hidden in the weakly populated LECLM "noise" is of interest. This will be reviewed in the Discussion section along with the question of artifacts which may exist in this analysis due to the intrinsic topology of the ECLM surface. To the author's knowledge no quantitative evidence of on-off intermittency in an actual biological system has ever been measured (although more general forms of intermittency have been investigated extensively).

### 2.3 Cluster Size Distribution

Fig. 4 is a log-log plot of probability distribution of the cluster sizes for all the synchronization clusters found for all LECLM. This distribution agrees quite well with the previously referenced coupled logistic map study [17] where dynamic criteria was used to establish new links in the evolving network and scale-free structure was found. They found a cluster distribution with a power-law part followed by an exponential cutoff (beginning near the end of this ECLM distribution) after 100000 iterations of their model. In their study they also extended the cluster distribution curve to model systems with ~500 elements. This result appears to again coincide with their connection criteria in actual evolving dynamical systems.

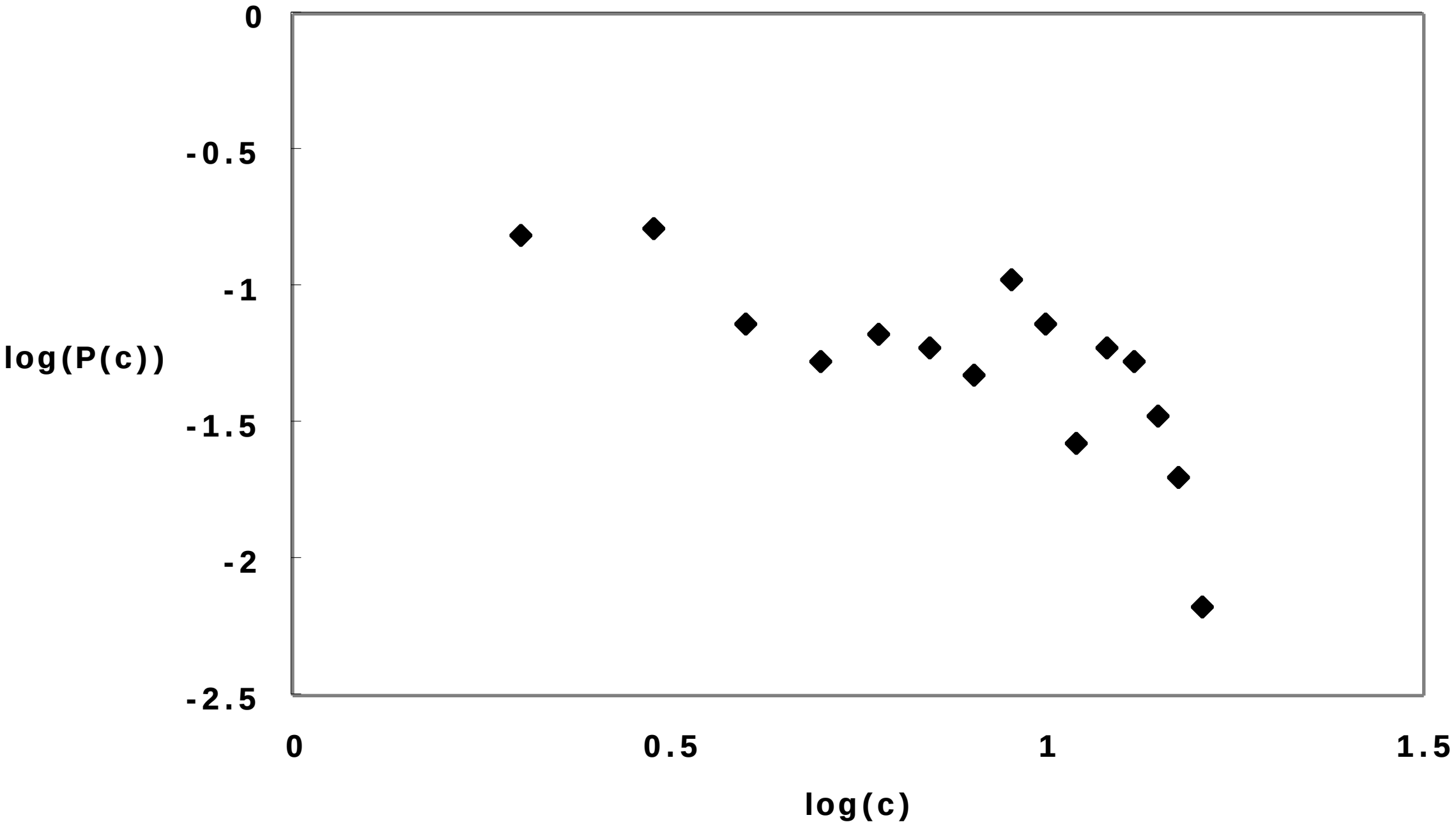

Fig. 4: Log-log probability distribution of **c** = cluster size.

### 2.4 Energy Dissipation

Since this is a dissipative open system, energy use and the relation to dynamics is an important consideration. One can interpret the number of LECLM as the number of ways available to dissipate energy into the system through "driving" or "driven" synchronization between genes. In this interpretation, if one rank orders the number of genes with at least one synchronization state (LECLM), two synchronization states (LECLM), up to 8 (the maximum # of LECLM, or links, any single gene in the experiment has), then this can be viewed as an energy dissipation distribution for a randomly selected gene within the system. This is based on the notion that more energy will be dissipated into the system by a gene that has a higher number of synchronization states [53]. Fig. 5 is a log-log plot of **P(M),** the probability of a given gene driving **M** synchronization modes. Once again the data sample is small but a clear power-law scaling is seen (scaling constant ~2 and linear correlation > .96). A recent study suggested that almost all biological systems use hierarchical fractal-like networks as a way to minimize energy dissipation by efficiently transporting it between spatial scales [33] this effect occurs in other natural systems as well [34]. One can speculate that this finding is a variant of this type of energy dissipation minimization principle in a genetic network [35]. The interpretation of Fig. 5 as an energy dissipation PDF also seems to favorably match characteristics for the stock market example (see "Energy Dissipation" in the following section).

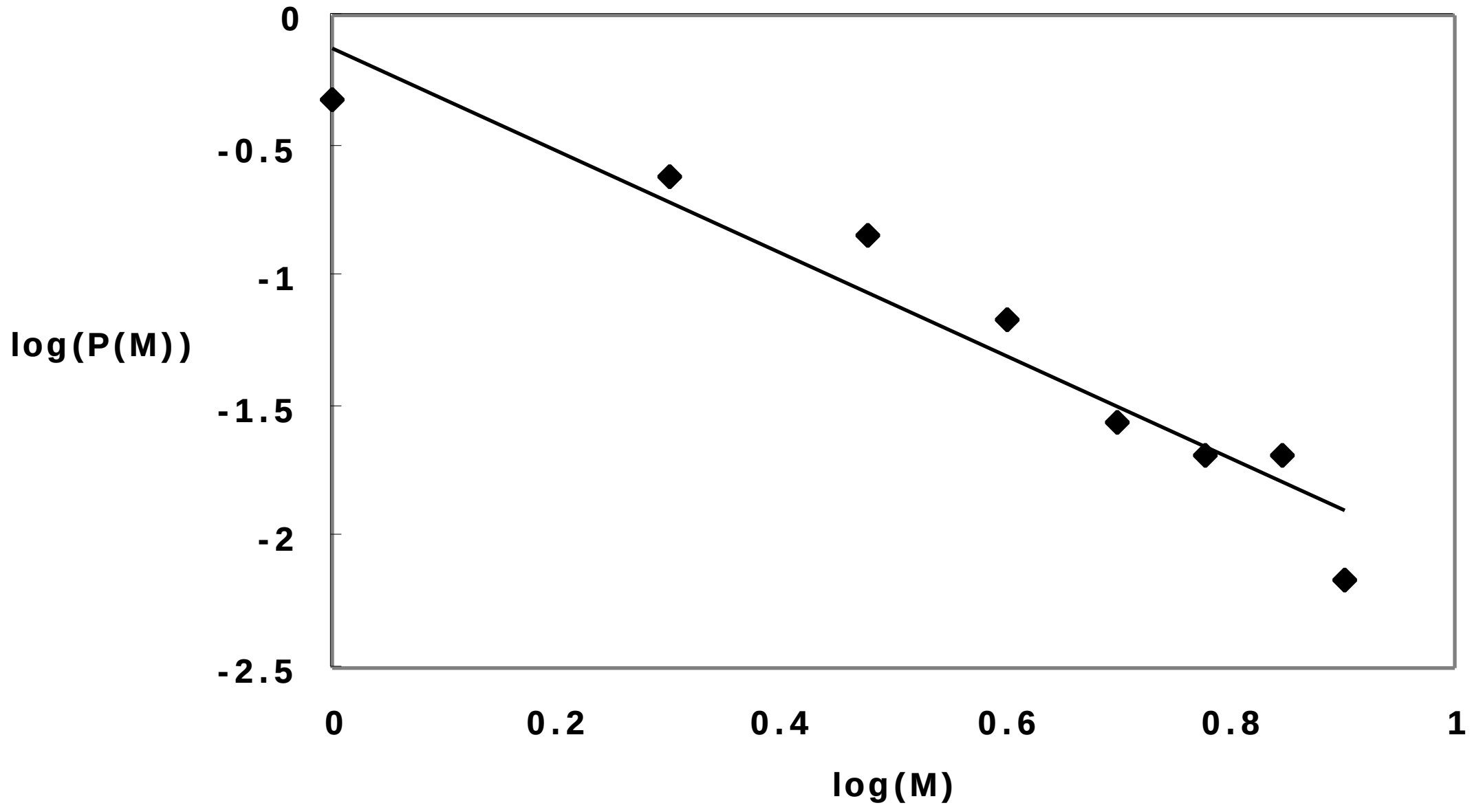

Fig. 5: Log-Log probability distribution of M = # of synchronization modes (energy states).

## 3. ECLM Analysis of a Stock Market Network

The input data for this ECLM trial consisted of ten price points for 26 stock market indices and two stocks. These price points represent the average of daily closing prices over two week intervals for a year (April 2001-April 2002). The ECLM methodology for this experiment is described in the Appendix. In summary, each of the 10 time points (including the 500 corresponding snapshots) were fitted to an ECLM point model with |**C**| > .95, where **C** is the Pearson correlation coefficient (assuming such a model could be found). All the fitted point models were then compared to each other to see which ones had pairwise |**C**| >.95 over the 10 (model) time points. It was found that the model network is fully connected, with every stock having at least one link to every other stock.

### 3.1 Connectivity Distribution

In Fig. 6 you see the log-log plot of **P(k)** for the connectivity of LECLM models as in (2.1). There are only a small number of points from which to analyze a pattern, but clearly this is a different situation than the connectivity in the genetic network. If anything, P(k) is more like a random distribution with a Gaussian or Poisson-like shape. The difference from the genetic network is not surprising. In the genetic network there were 146 synchronization modes (LECLM) between 69 genes. The largest synchronization cluster was 16 out of the original 112 genes [58]. In the stock market network, there is at least one synchronization mode that connects every gene to every other gene. Also, there are only 18 different modes for all 28 stocks indicators, and there are two dominant clusters that were found to contain almost all the stocks [36]. The apparently random distribution in this setting seems to reflect coupled lattice models where the system in a turbulent (highly chaotic) state with strong global coupling and clusters of synchronization, even though the local coupling is weak [37-41]. This picture is also consistent with the fact that there were no bounded LECLM found in the ECLM analysis of the data. Since we are analyzing stock market indices (except for two stocks) reflecting a group of stocks, and often in markets a "rising tide tends to lift all boats" (or wrecks them!), it is not surprising to see strong global behavior dominate over index to index (local) behavior. The random connectivity distribution with this type of global behavior is indicative of a randomly connected small-world SW network [42]. (SW networks can also be scale-free).

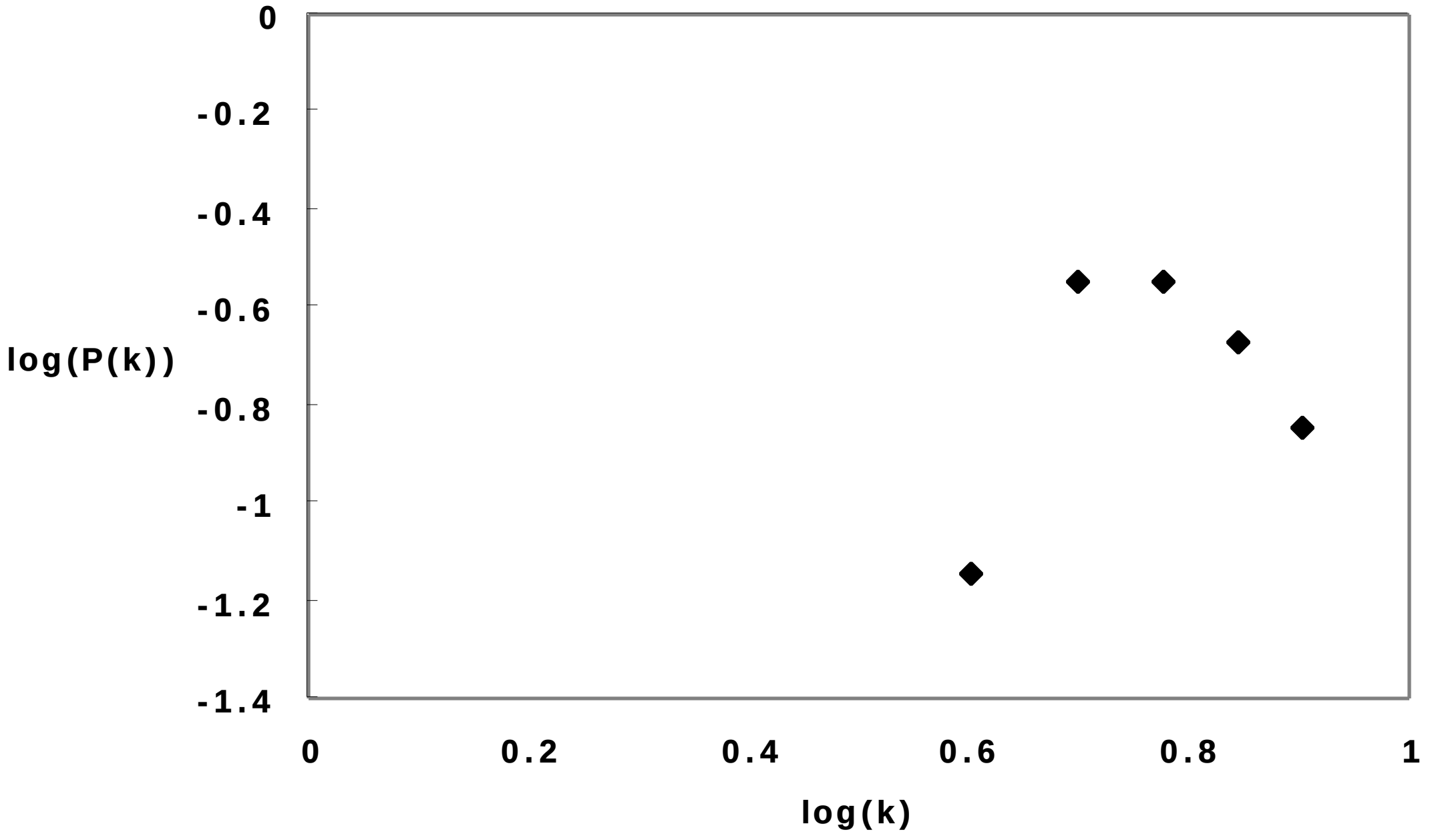

Fig. 6: Log-Log probability distribution for linked ECLM models derived from stock market data.

### 3.2 On-Off Intermittency

Fig. 7 is the analogous plot to Fig. 2 for the gene network. The scaling indicated by the line is ~2. As in (2.2) we remove the weakly populated states with only 1 or 2 values for Fig. 8. In Fig. 8 the slope of the line is ~3 with a linear correlation of > .95. If this is another indication of on-off intermittency, it coincides with transient synchronization states found in the coupled models referenced in (3.1) [17]. On-off intermittency has been previously been measured in financial market data [31].

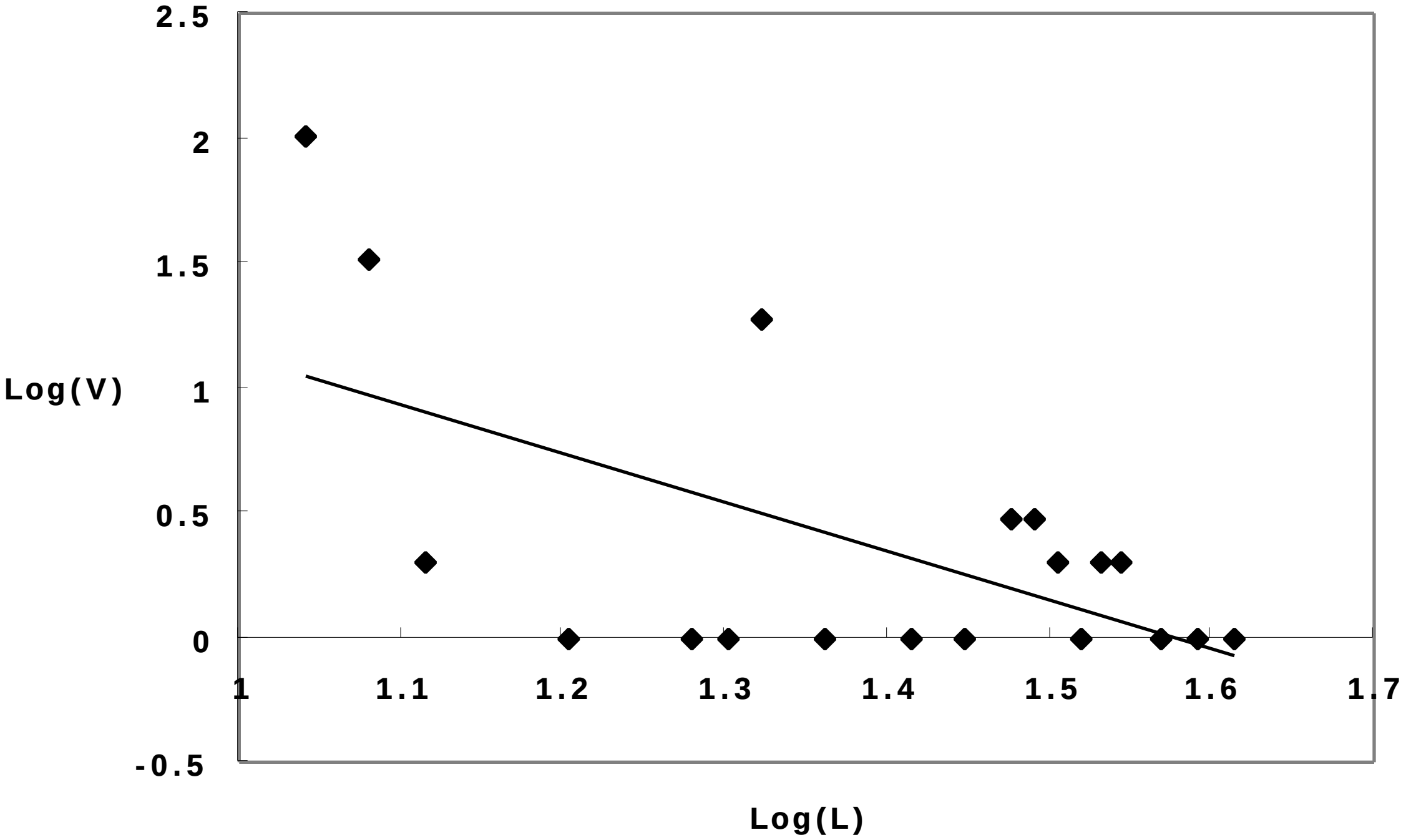

Fig. 7: Log-Log distribution where **V** = the unbounded LECLM cutoff iteration values and **L** = the number of unbounded LECLM with value **V**

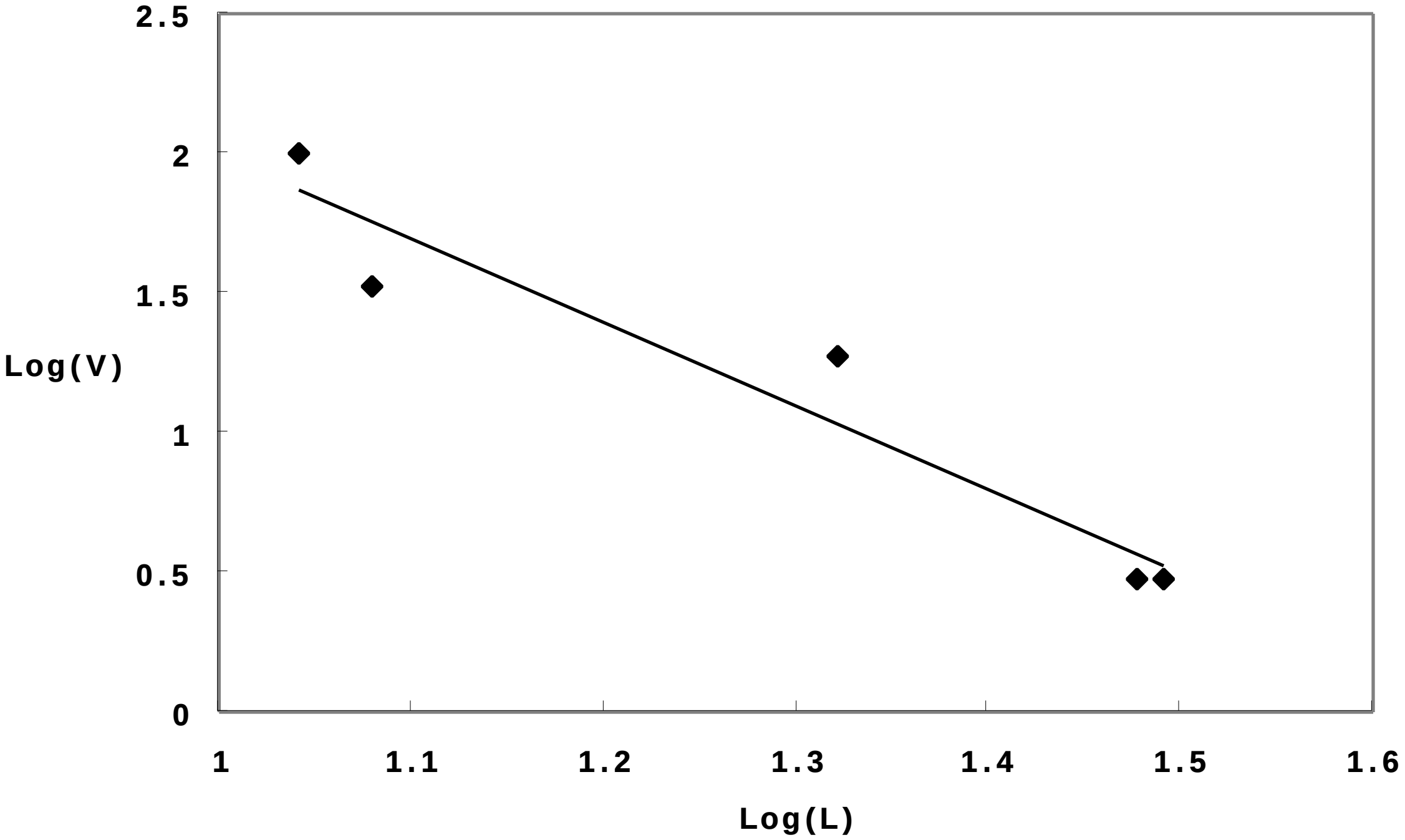

Fig. 8: Same as Fig. 7 with weakly populated value states removed.

It is of interest that the scaling is so close to 3 in Fig. 8. This is due to the fact that "universal" period 3 behavior has been found in globally coupled logistic maps exhibiting turbulence and loosely coupled (locally) behavior [43]. If such phenomena is connected to real world situations this could possibly act as guide to directly connect ECLM locations on the complex plane (those with period three behavior) with a particular type of complex system behavior (See Appendix: "Mappings"). Many of the LECLM clusters associated with analysis show period-3 type behavior (see "LECLM analysis") as well as generally low period behavior consistent with the coupled map studies of this type of system behavior [43]. This again brings up the question of artifacts due to the intrinsic topology of the ECLM surface, which will be reviewed in the Discussion section. The wide dispersion of LECLM clusters for the stock data (as opposed to the genetic data) seems to indicate behavior which is more "bursty" (see Appendix) than the genetic data. This bursty behavior is consistent with other studies of financial data [44].

### 3.3 Cluster Size Distribution

Fig, 9 is the probability distribution of the cluster size for all the synchronization clusters found for all LECLM in the stock market data. There is does not appear to be much structure in the point scatter. In the LECLM there are two clusters that each contain almost all the points. If one interprets this situation as a network that is near or has crossed the percolation threshold [45], than the cluster distribution will be diffusive and non-stationary [46]. This also coincides with the random but synchronized picture that appears to be present. An ECLM analysis of stock data from the previous year, showed a different (an equally scattered) distribution [36].

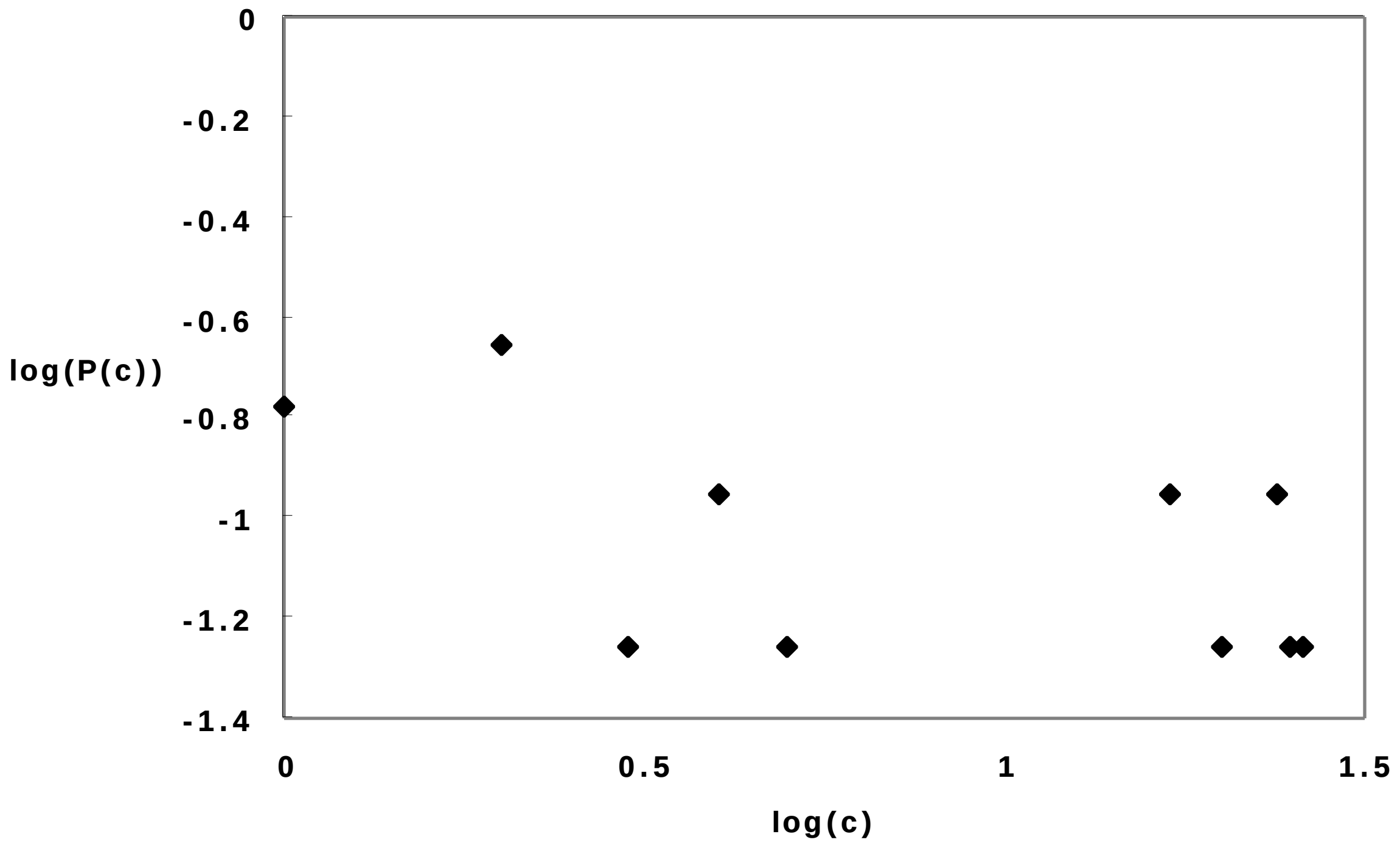

Fig. 9: Log-log probability distribution of **c** = cluster size.

### 3.4 Energy Dissipation

In the manner of (2.4) Fig 10 is the log-log plot of the energy dissipation PDF analogous to Fig, 5. The lognormal-like shape is consistent with the energy dissipation PDF of a multfractal cascade of the kind found in models of turbulence [49]. This interpretation fits with the turbulent picture previously outlined and also gives validation to interpreting this state distribution as an energy dissipation PDF for a randomly selected gene. In multifractal turbulence, the energy is spread by a hierarchical cascade through progressively finer scales and dissipated as heat [47]. This is to be contrasted with efficient monofractal (one scaling constant) minimized energy dissipation discussed in (2.4). Multifractal and turbulent behavior in financial data has been a topic of much study [44, 48].

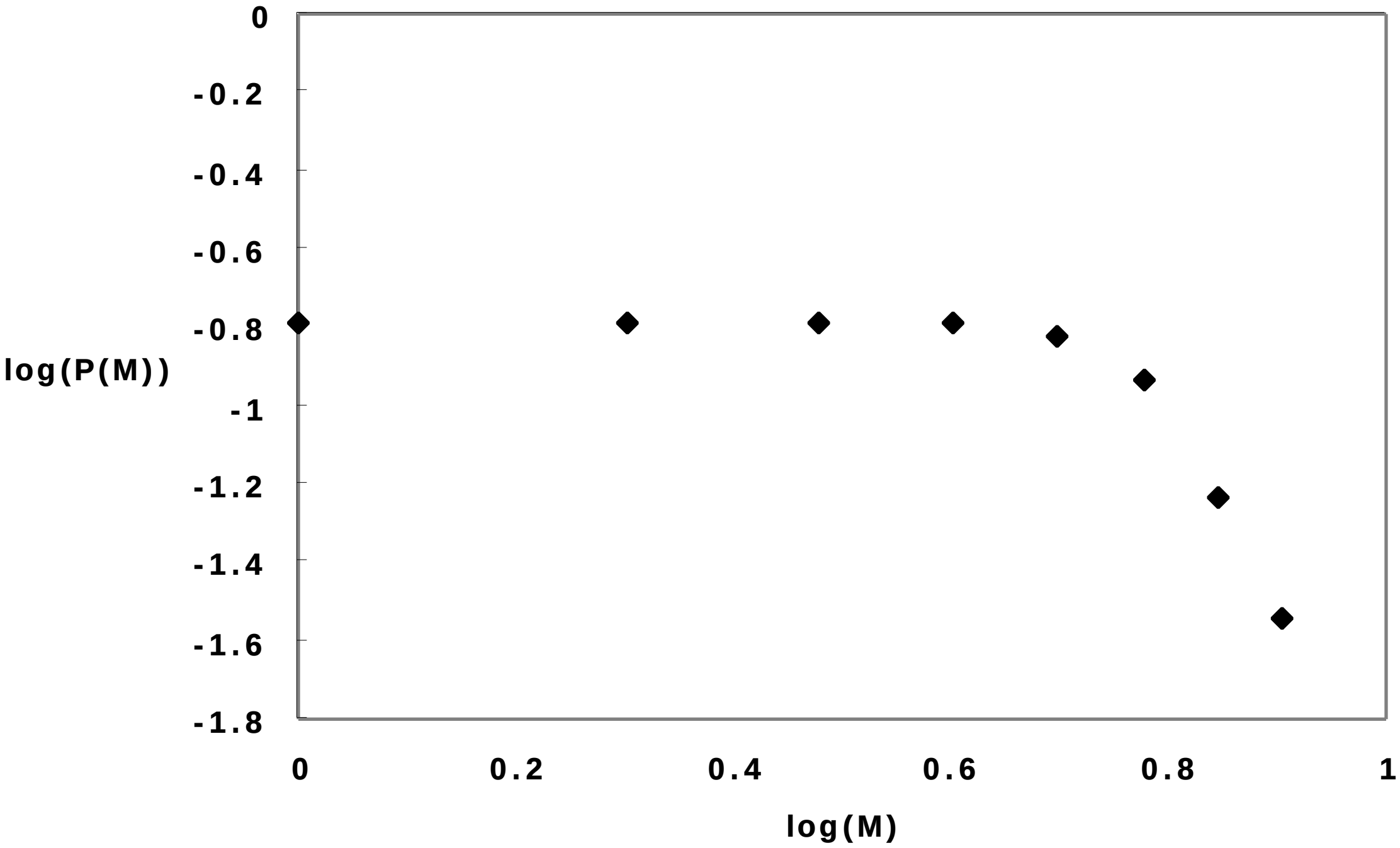

Fig 10: Log-Log probability distribution of M = # of synchronization modes (energy states).

## 4. LECLM Analysis

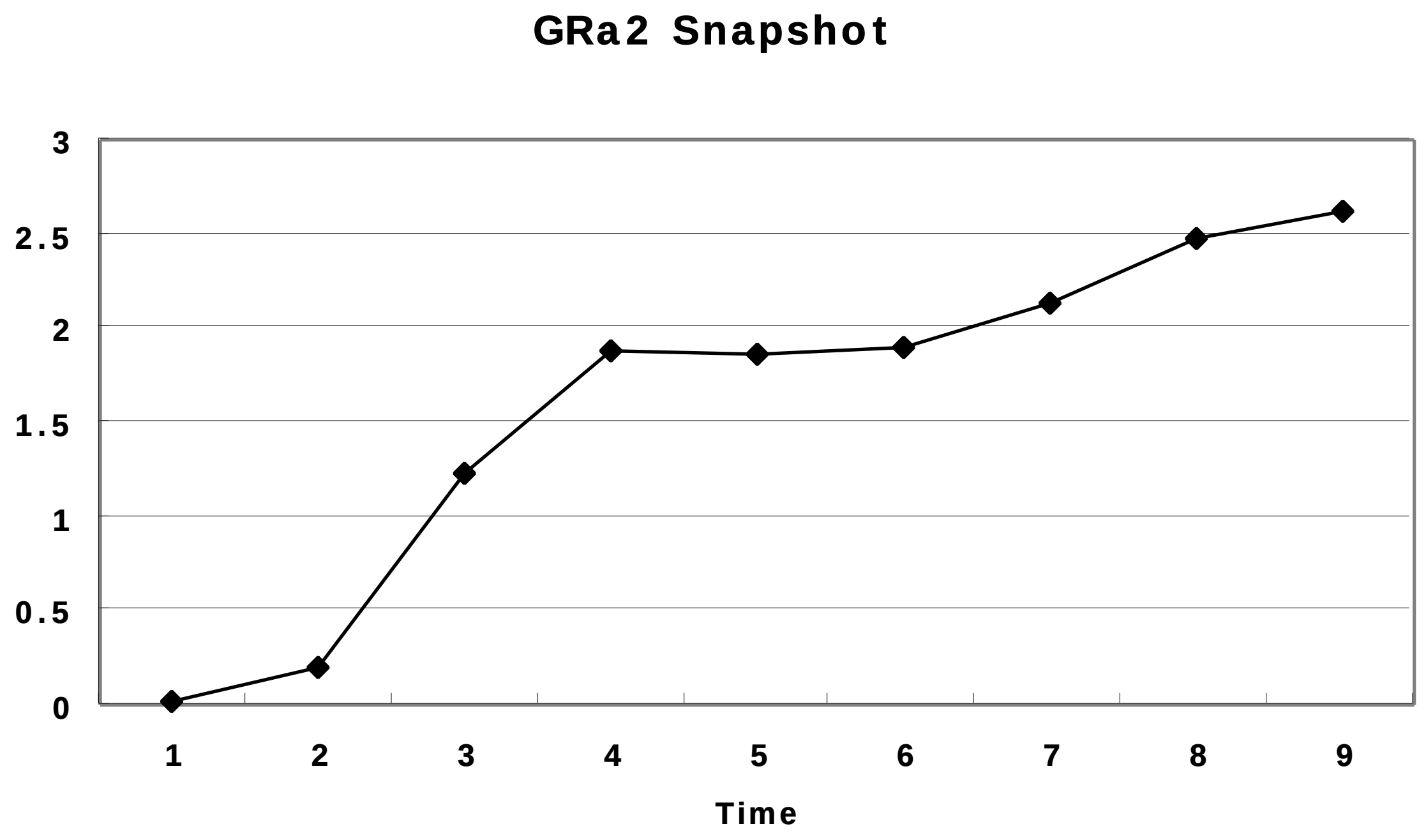

We briefly introduce the ECLM analogy to wavelet analysis, LECLM analysis. Shown below and the next page are the plots of two correlated gene LECLM, representing a synchronization mode between them. Note the small difference between the graphs for the first nine time points. Subtle differences in behavior can be noted in behavior at similar time points (as in wavelet analysis [4]). The original GRa2 snapshot (see "Preprocessing the Data" in the Appendix) from which the correlated GRa2 LECLM was derived is shown above for reference.

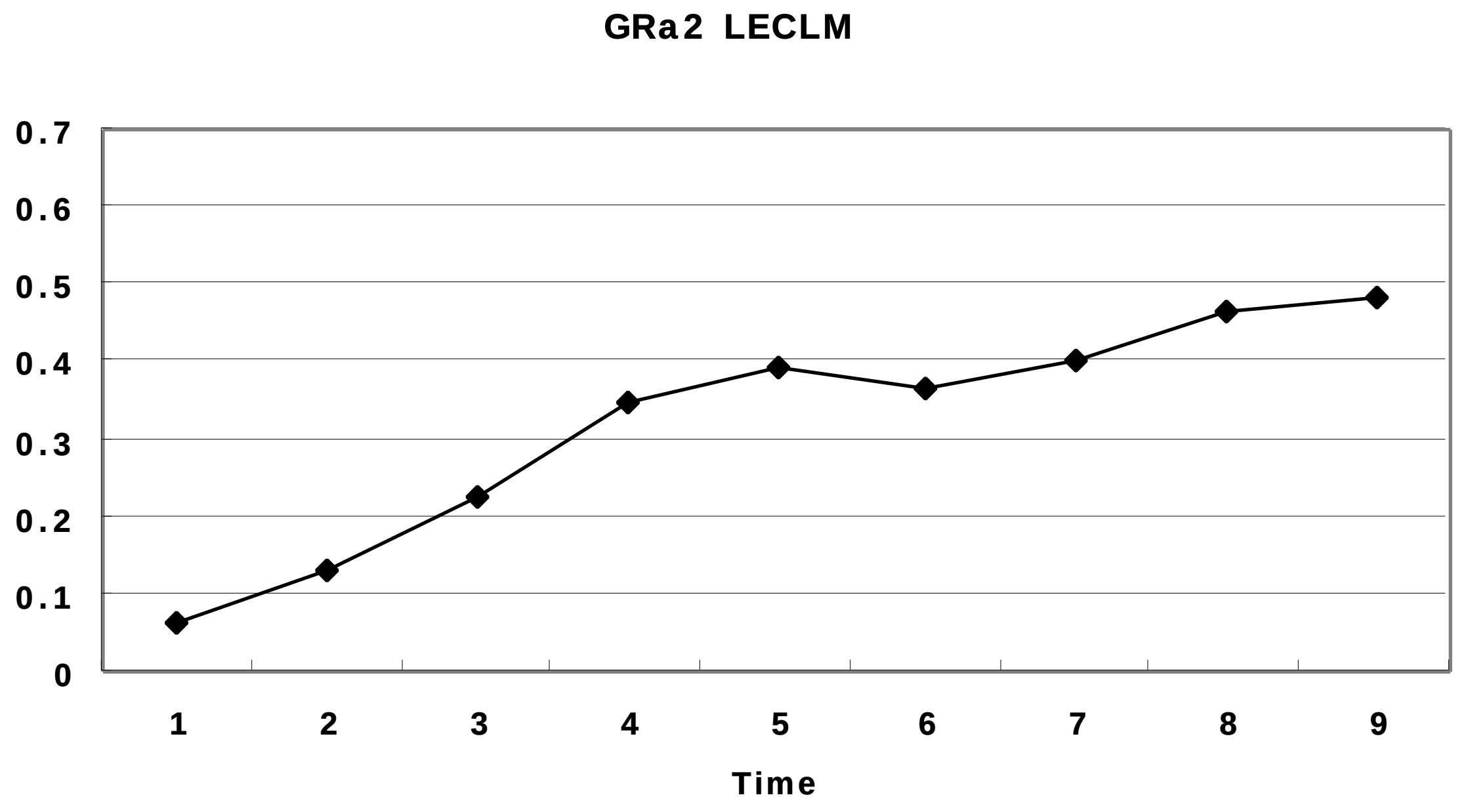

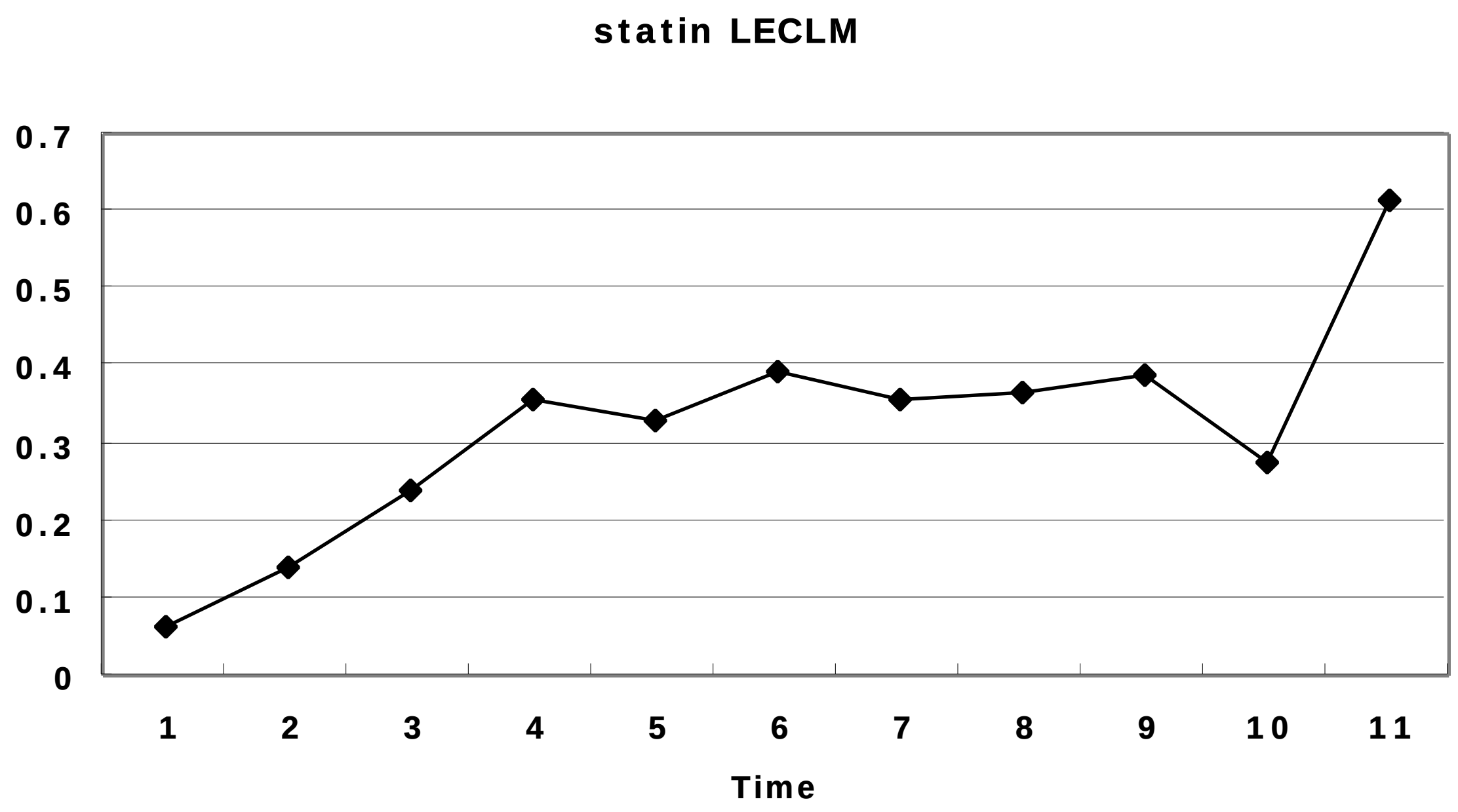

Note that there are extra time points in the statin plot due to a longer bounded state for this LECLM. LECLM can be used for predictive analysis. The stock LECLM, shown below and the next page, are correlated for the first ten points (days) but both are bounded for a longer period and one is bounded for a much longer period (20 days). Such predictive capability could be very powerful in a network setting when applied at hubs, particularly driving hubs (see "Network Graphs" below). LECLM analysis illustrates how dynamic simulations of real-world complex network environments could be constructed as well.

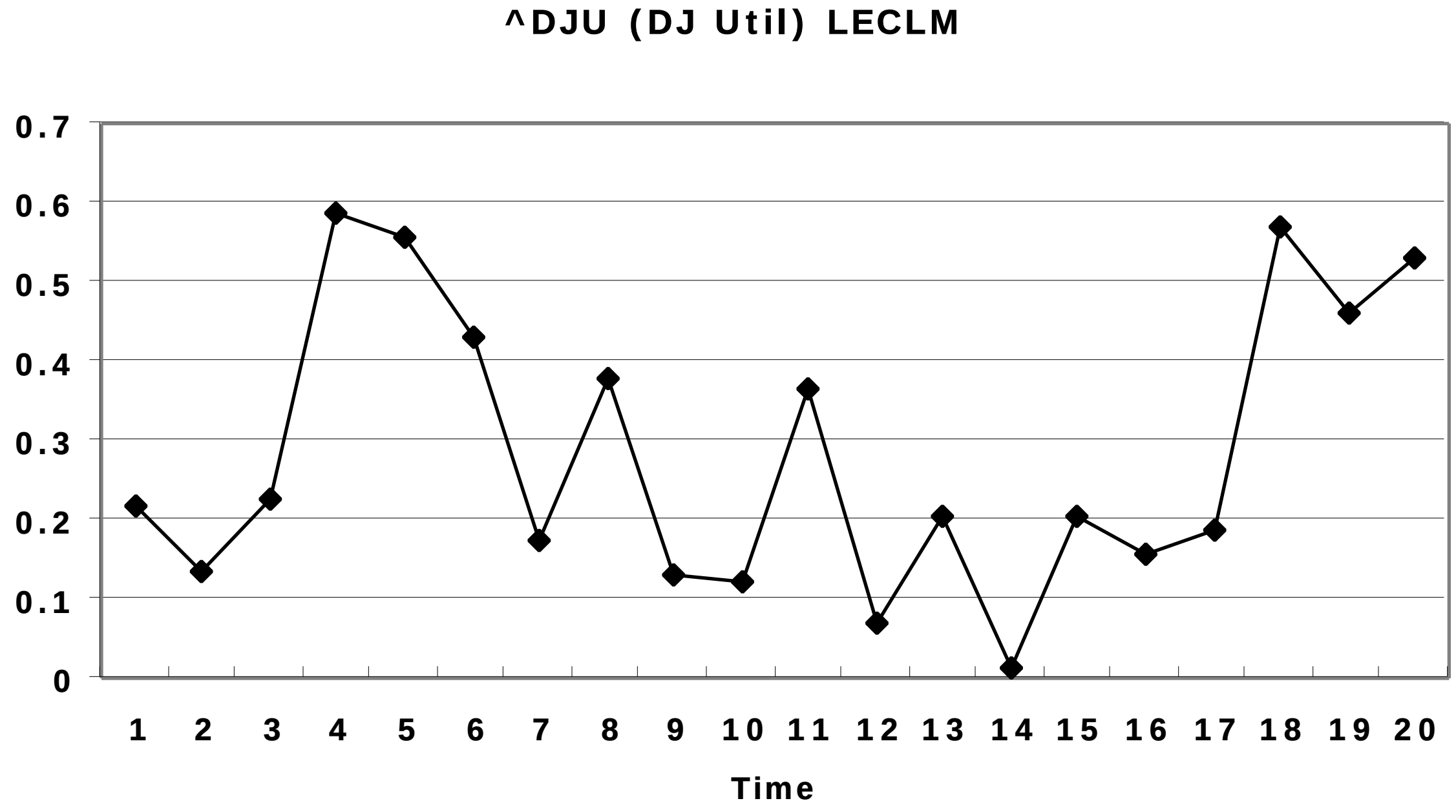

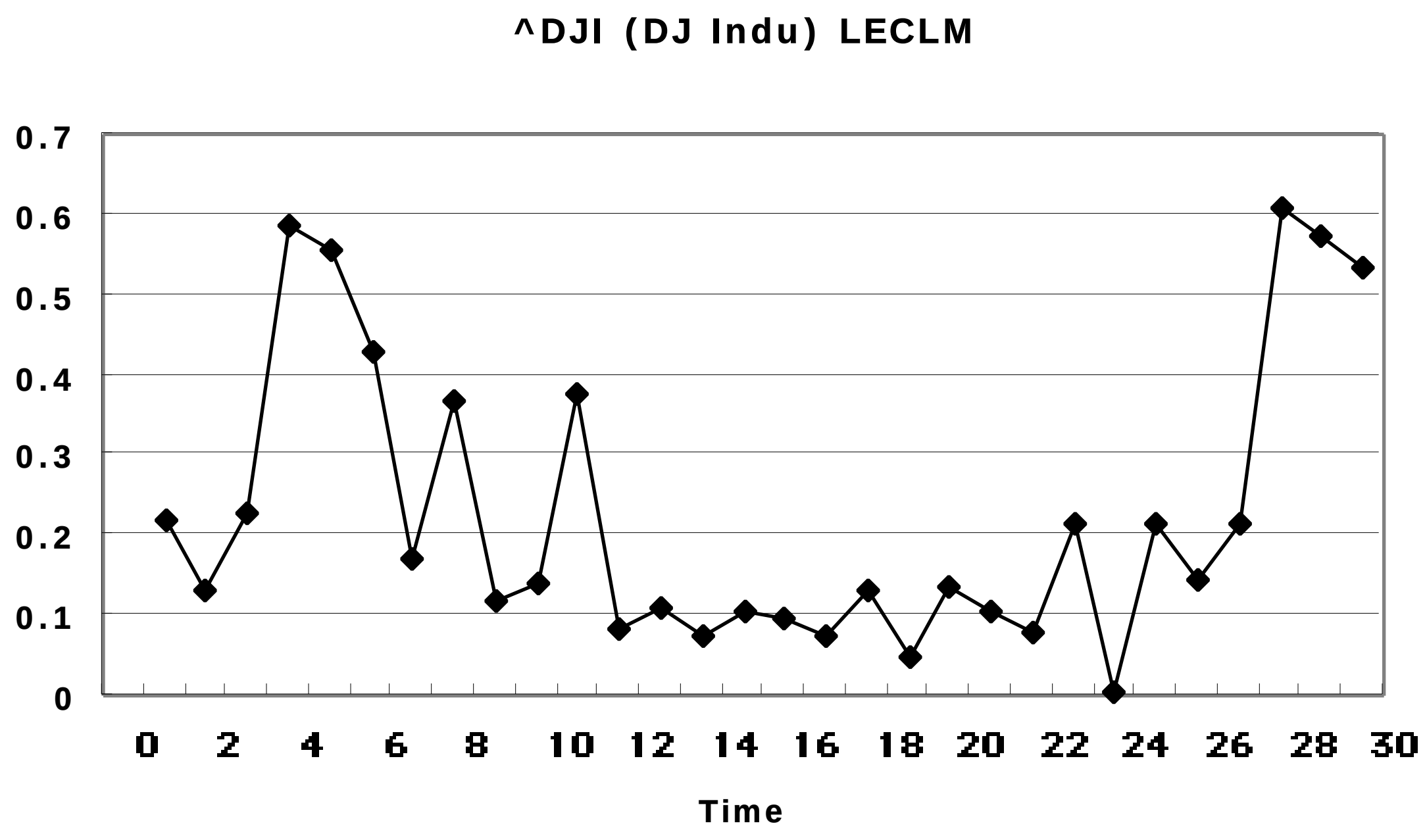

## 5. Network Graphs

Any method which allows network analysis of data is open to graph theoretic analysis. ECLM is no exception. The network graph of all LECLM for the largest gene network of 59 is complex and large [51]. Rather than deal with the complexity of such a graph. It can be broken into subgraphs of interest. The figure on page 14 is made up of the first order (direct) direct links to the GAD 65 gene, an important gene in the network [50].

One can also create a directed graph as seen on page 15. This could be called a "driving" diagram of the subnetwork. The genes where the arrows originate have a greater number of synchronization modes **M** (see 2.4) which are NOT used in linking to the gene than the gene they are linking (pointing) to. This is interpreted as the originating gene having more dynamical connections outside this particular link, and to therefore be more likely to drive the interaction (see 3.1). From the standpoint of attractor dynamics one can also view this as the originating gene providing more escape routes for the dynamics from one cluster to another [53]. Lines with no arrow represents links where each gene is equally likely to drive the interaction. The figure following the undirected graph is a driving diagram of the first and second order genes connected to GAD65 (all its direct links and all their direct links). The distribution of driving links (outgoing arrows) in this diagram follows a power-law distribution [54]. This is especially interesting in the context of efficient energy dissipation expressed in (2.4) [35]. The driven links (incoming areas) tend to follow a power-law distribution also, as can be seen in the "clumping" of arrows around certain genes in the diagram. This would make energy transfer, as well as direct or indirect biochemical transfer, between the driving and driven hubs very robust under random failure of links within the network [52]. This robustness has been suggested as a reason for scale-free networks in some natural systems [75]. Similar methods can also be used to graph the findings from the stock market network [36].

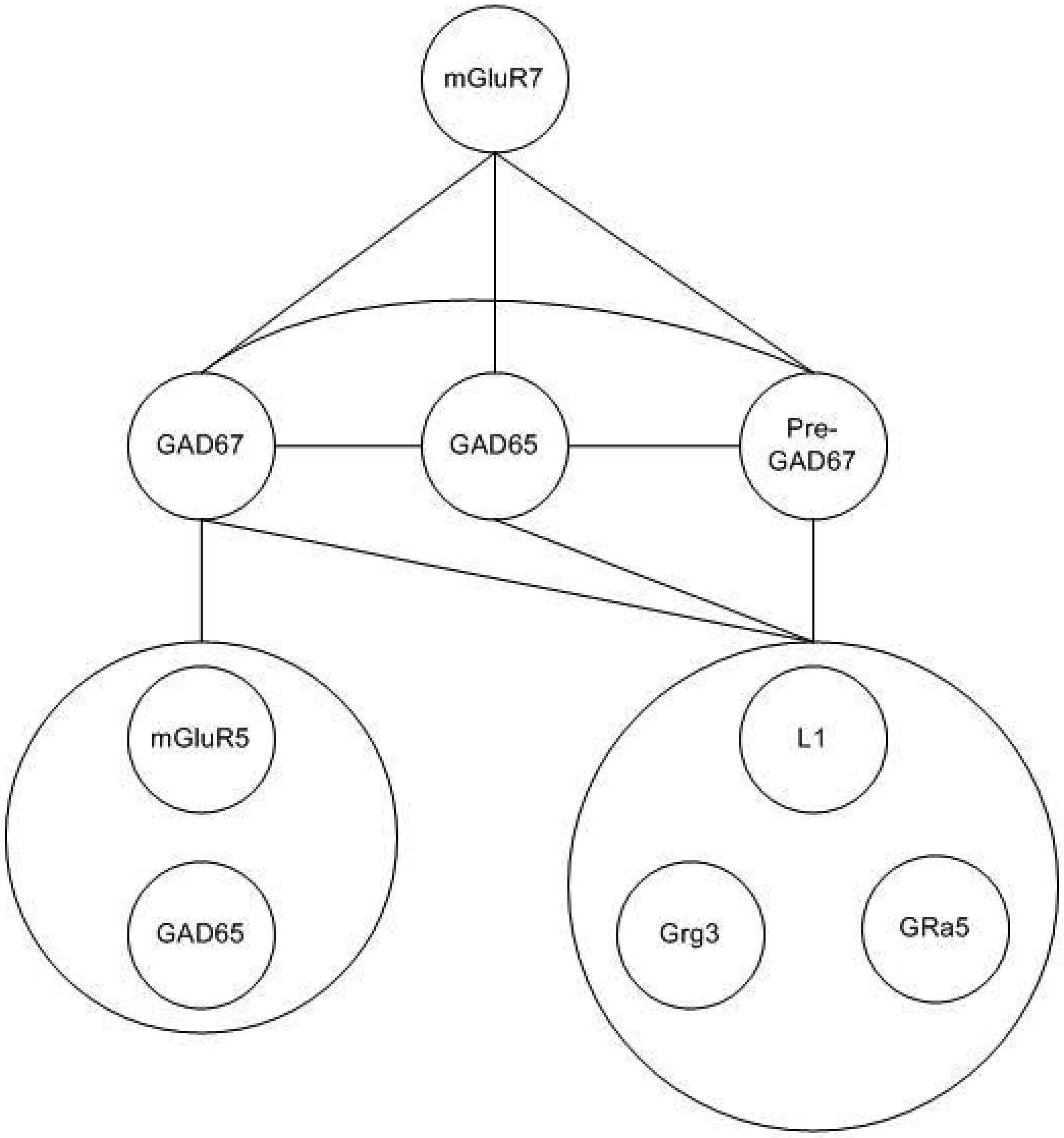

**First-Order GAD65 Subnetwork**

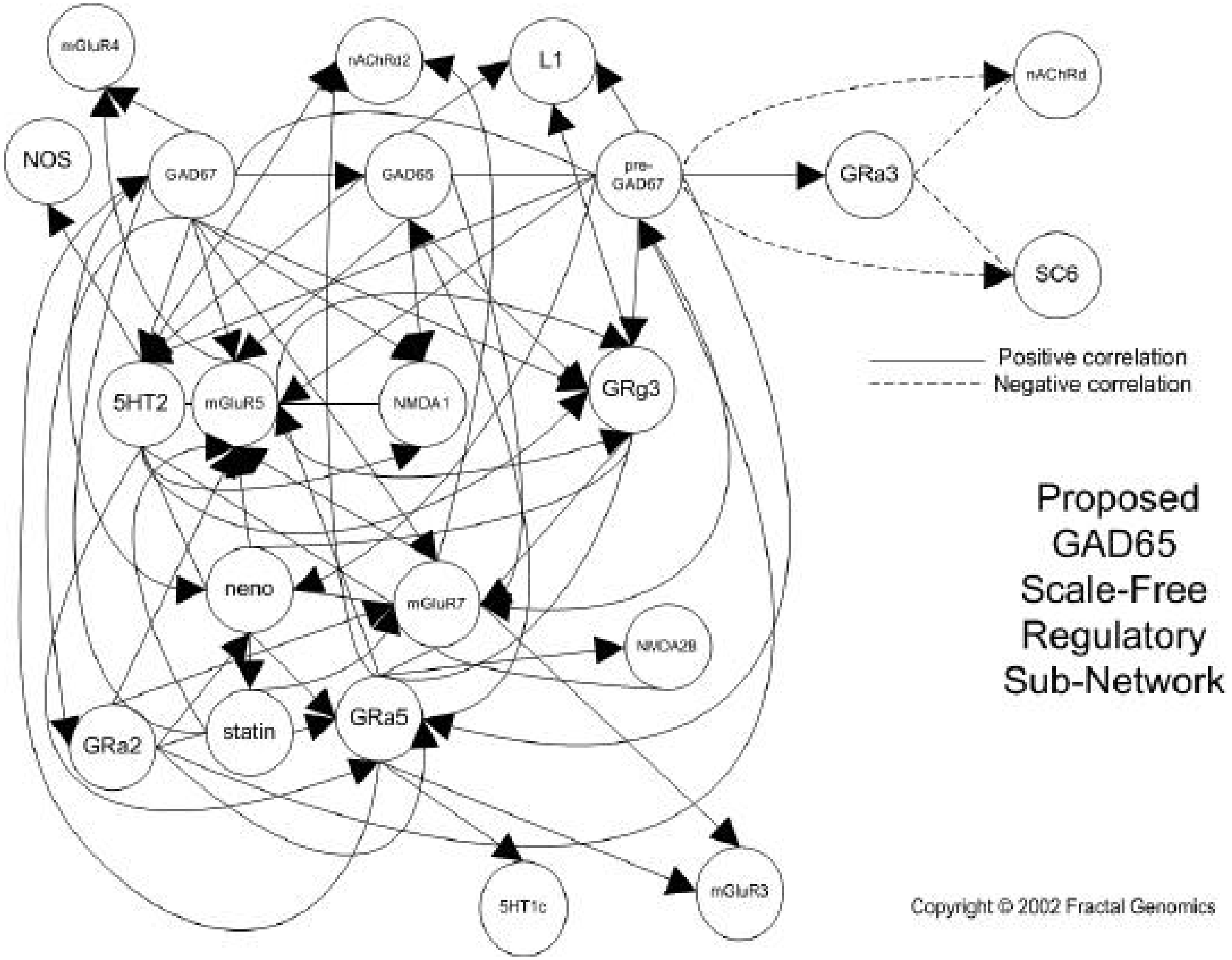

## 6. Discussion

Before discussing the results, questions regarding artifacts of method need to be addressed. That is, how much does the topology of Mandlebrot Set and the topology near its boundary impact the output of ECLM? Are the examples studied in this paper a reflection of the dynamics intrinsic to the complex logistic map or are the dynamics of the complex logistic map somehow a reflection of the underlying dynamics of the examples? The successes displayed by coupled logistic map models would tend to lead one to the latter conclusion. Also, many of the results found in this study seem to make sense within the environment of the complex systems studied and agree with other experiments unrelated to chaotic dynamics. The scale-free structure found in other biological systems is a good example [25-28]. The fact that ECLM seemed to show different behavior between the genetic network and the financial data is another indicator that something is truly being modeled other than the topology of the surface itself. ECLM analysis of more examples will help bear this out. Also, further probes into the data produced from this study need to be done. The genetic network results need to be more fully investigated from a biological perspective. The genetic network study produced a vast amount of information that is only just beginning to be analyzed [35]. This is one direction of future study along with general validation of ECLM.

The question of on-off intermittency needs to be addressed also. It has been assumed that if ECLM is accurately modeling the dynamics of these examples, then it is really probing long-range order in on-off intermittency. This is because in on-off intermittency, not only do the synchronization states show scaling behavior, but also the *amplitudes* of the variables involved [56]. It is possible that some other type of scaling behavior in the synchronization states is being probed. It could be that there are "superpersistent" states that are appearing [55]. On-off intermittency occurs in the glassy phase but the stock market example was in a "hidden coherence"[43] turbulent environment. Larger datasets will be studied to investigate what it is that ECLM is actually probing and attempt to find some fundamental relationship between the source data and the scaling exponents [56]. It is possible that a snapshot of on-off intermittency is being seen, perhaps with additional scaling overlaid on it [32]. More studies may also reveal that some of the findings in the genetic network analysis are unique to the neurochemistry of the particular network studied [58].

The question of the impact of the bounded LECLM on the overall dynamics of the genetic network was largely ignored. These states could have a profound effect on the evolving dynamics of the system [57]. This phenomenon also needs to be explored in ongoing ECLM research.

Summarizing the result found in this study, there appears to be new results involving scale-free structure, on-off intermittency, and energy dissipation in biological systems. Similar methods and interpretations seem to yield consistent and intriguing results in the vastly different world of financial systems. These results, if accurate, seems to lend more weight to the notions of universal behavior in complex systems. Because ECLM is data independent many other types of systems are open to analysis. This makes "universality" testing possible across a variety of complex systems. Another benefit stemming from ECLM data independence is the ability to analyze heterogeneous data. For example, genetic data, cellular data, and clinical data could be analyzed all on the same complex map. This could make it possible to perform multiscale analysis of biological systems within a single framework, one of the major goals in the relatively new field of systems biology [59].

LECLM analysis could potentially represent a field of analysis in itself. Powerful simulations with predictive power could be studied that have evolving dynamics and topology based on actual data. The predictive power of LECLM analysis will be examined further in the process of ECLM validation and research.

Real-time ECLM analysis has yet to be explored in depth. This is accomplished by replacing the snapshot attractors with real-time snapshots (see "Preprocessing the Data" in the Appendix). ECLM analysis of single sequences can be performed rapidly and ECLM analysis of datasets can be done incrementally, as the data appears. Real-time dynamics could look very different than what is seen in static analysis. This is another area of future study.

These preliminary results appear to produce some insight into what may have been previously hidden dynamics. Holomorphic Dynamics is the dynamics induced by the iteration of various analytic maps in complex number spaces [60]. Holomorphic Dynamics moves closer to physical world dynamics if ECLM proves a valuable tool. Perhaps, buried within the complex plane, there is more physical meaning than we realize.

# Appendix: ECLM Methodology

## I. Generating ECLM Models

The values to be fitted by the time series data are generated using the generating function for the complex logistic map:

$$z_{n+1} = z_n^2 + c$$

Where **c** is a point in the complex plane and $z_0 = 0$. The function is iterated **N** times where **N** = the number of points in the data sample. Each of the $z_n$ generated is normalized to a real number between 0 and 1[70]. This number is then multiplied by the difference between the upper and lower bound of the measurement (in the examples used this is twice the standard deviation) and added to the lower bound of the measurement.

This operation is performed for each of the **N** iterations. The Pearson correlation coefficient **C** is calculated between the model sequence and the original data. Another point **c** is chosen and the process is repeated again. In the actual ECLM process, this point was chosen progressively on a 700 by 700 grid moving in the x (real) direction. This grid covered and area slightly larger than the area covered by the Mandelbrot Set. Points where $|z_n|$ reached a value greater than 2 were disregarded.

This process was repeated for every point on the grid. The point corresponding to the model with the greatest value of **C** was used as the center point for another grid with the same number of points but a smaller size. The modeling process was then repeating for this grid (analogous to producing a Mandelbrot "zoom"[11]).

The zooming and modeling process is repeated until the value of **C** produced is less than or equal than the previous zoom. This process has been shown to converge efficiently to "best fit" models for given scoring criteria [64]. If the best fit $|C| > .95$ then it is assigned to this data sample.

## II. Linking the Models

Beginning with clusters of point models that are within some threshold (Euclidean) distance of each other, the **N** points assigned to each point are (Pearson) correlated against each other pair by pair. If $|C| > .95$ then the **N** points assigned to each model are placed in a table to be correlated against every other unique pair of **N** points found which also fit this criteria. In the examples in this paper this process was terminated for each cluster when no further points could be found within the threshold distance of the edge of the cluster. Generally, these models tend to be dissimilar. The figure below illustrates this effect. Sequences with known similarities and differences were mapped onto the complex plane using an earlier variant of the ECLM process [64]. This plot indicates the "threshold effect" of a distance above which significant similarity (# of matches out of 1250) was unlikely to be found.

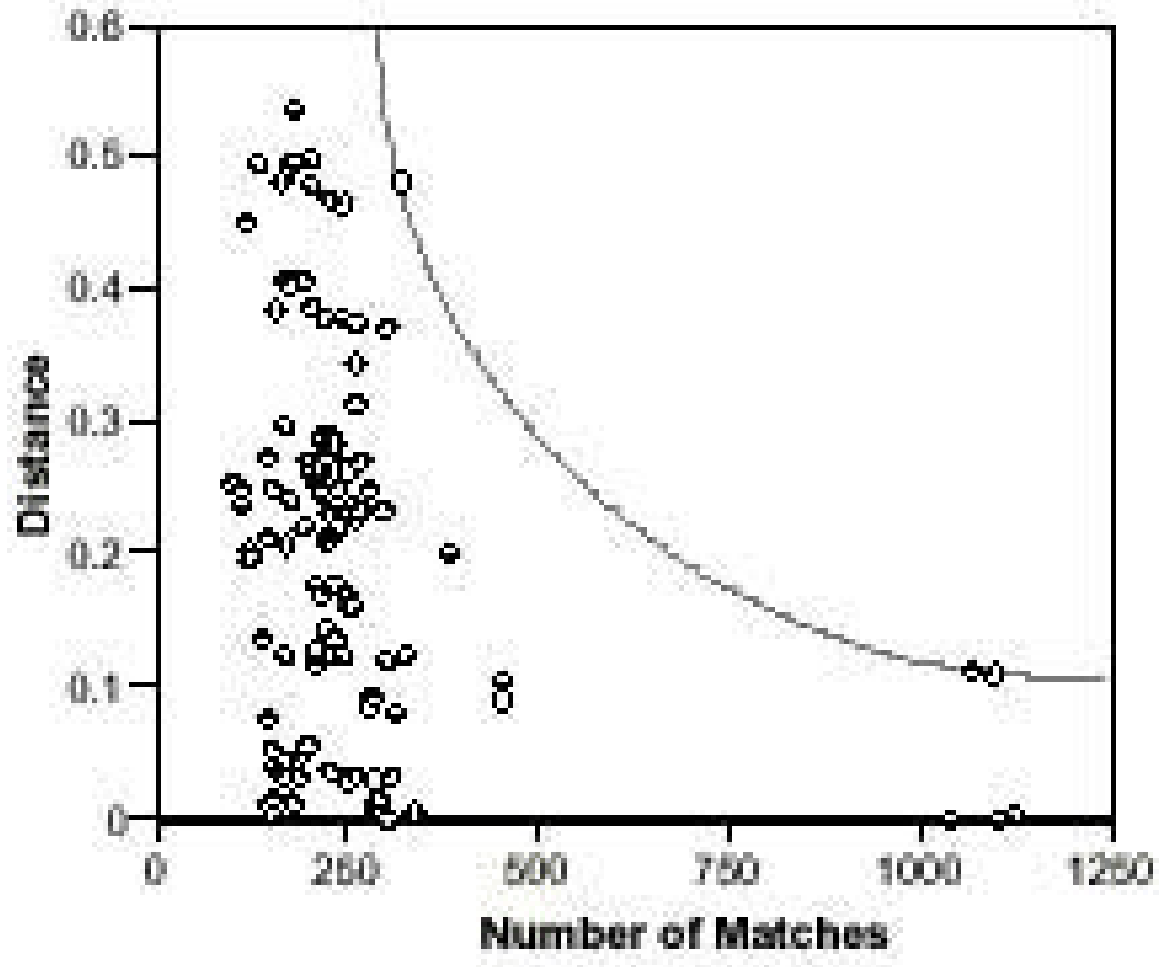

The process is continued for all clusters which meet the threshold criteria, progressively building the model table as the process continues. When all such clusters on the plane have been processed the procedure is terminated. All groupings of ECLM point models whose corresponding **N** points values showed $|\mathbf{C}| > .95$ with each other are recorded. The points in these groupings represent the linked embedded complex logistic maps (LECLM). This is the output of the ECLM process.

## III. Preprocessing the Data

The method of snapshot attractors [72] has been applied to finding the fractal dimension of chaotic attractors by looking at a "cloud" of trajectories starting from uniformly distributed initial conditions [61]. The idea of snapshot attractors has also been used in laboratory experiments to visualize and investigate fractal patterns arising in physical situations such as passive particles convected on the surface of fluid [62] and noise-induced enhancement of chemical reactions [76]. Snapshot attractors have also been utilized to study the transition to chaos in quasiperiodically driven dynamical systems [63]. In ECLM we apply this concept in a similar way.

At each data point in the original time series a uniformly distributed random number is added to the measurement in a manner analogous to the method used to form the iterated model values (falling between a standard deviation below and above the measurement). This is done a number of times for each time series, with ECLM performed on each modified time series. This inverts the snapshot attractor approach. Instead of finding the fractal dimension of the attractor in the ensemble of trajectories, we look for a fractal attractor that fits a single trajectory. It is interesting to note that the only (previous) physical meaning for Julia Sets that the author is aware of was a stroboscopic view (a series of snapshots) of kicked classical particle dynamics in a double-well potential [65].

## IV. Example Parameters

For the genetic network, the standard deviations used were supplied for each point with the data [16]. In the case of the stock market example, the standard deviations were calculated from the list of values for the average of the same day in a two-week interval for all two-week periods (ten values for each stock). ECLM was run on 100 preprocessed (snapshot attractor) series for each gene. ECLM was run on 500 preprocessed series for the each stock index. The stock indices used were ^DJT, ^DJT, ^DJU, ^DJA, ^NYA, ^IXIC, ^GSPC, ^OEX, ^MID, ^SML, ^IIX, ^IXY2, ^XMI, ^PSE, ^SOXX, ^RUI, ^RUT, ^RUA, ^DOT, ^VLIC, ^TYX, ^TNX, ^FVX, ^IRX, ^DJC, ^XAU, ^NDX (see list at http://finance.yahoo.com/m1?u) and two stocks, Intel (INTC) and HPLA Technologies (HPLA:PK) [73].

## V. Mappings

Below are the top-level view of the ECLM mappings for the stock data and a subset of the gene data. Note the tightly clustered dispersed form of the stock market data. This is consistent with a small frequency spectrum among the LECLM. The spread of points of the genetic data in the section near the back edge of the Mandelbrot Set, indicates a much wider and higher frequency spectrum [11, 66]. Interestingly, a similar pattern, on a different part of the map, was seen when an earlier version of ECLM was used to map Expressed Sequence Tags (ESTs) in DNA [67].

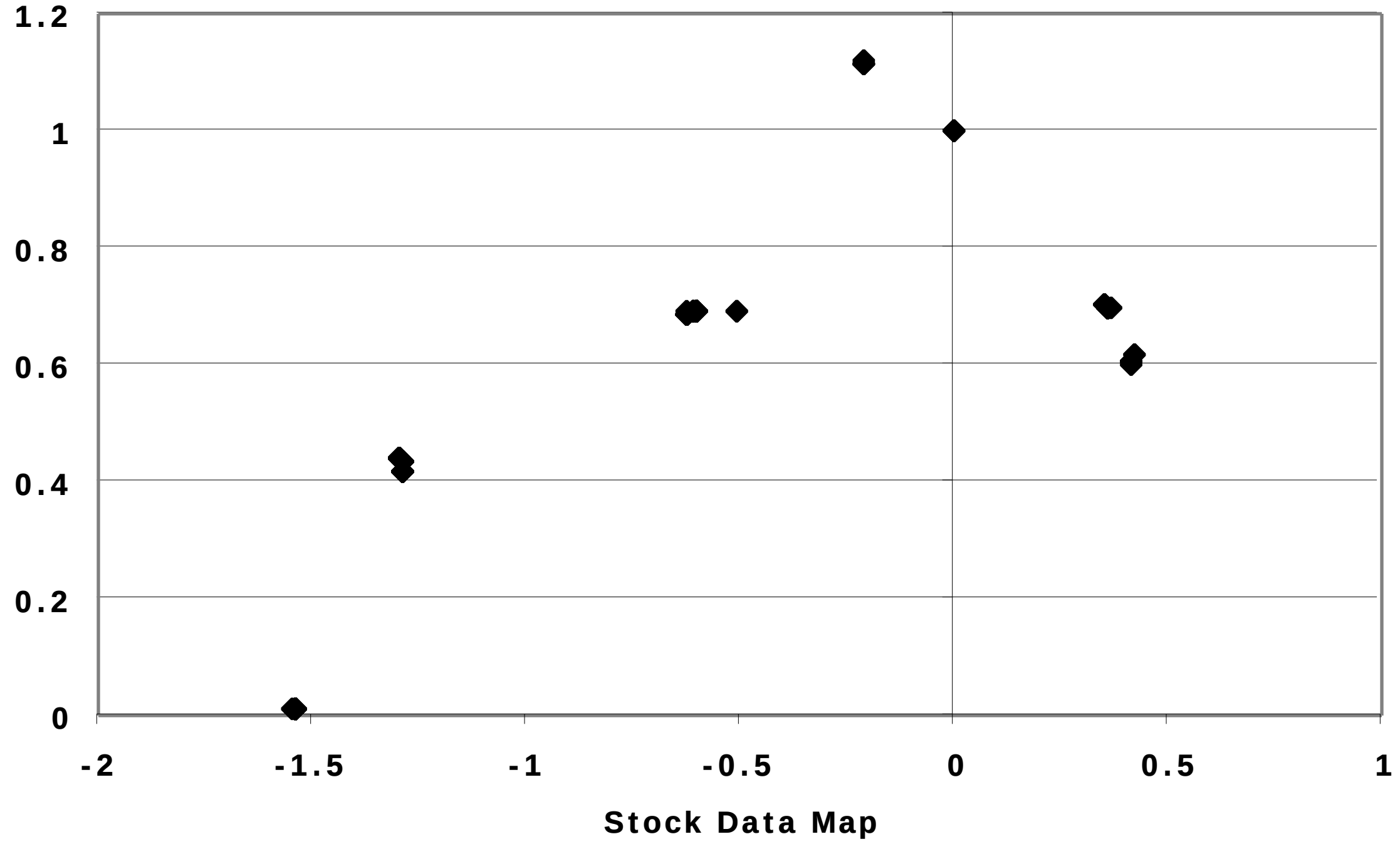

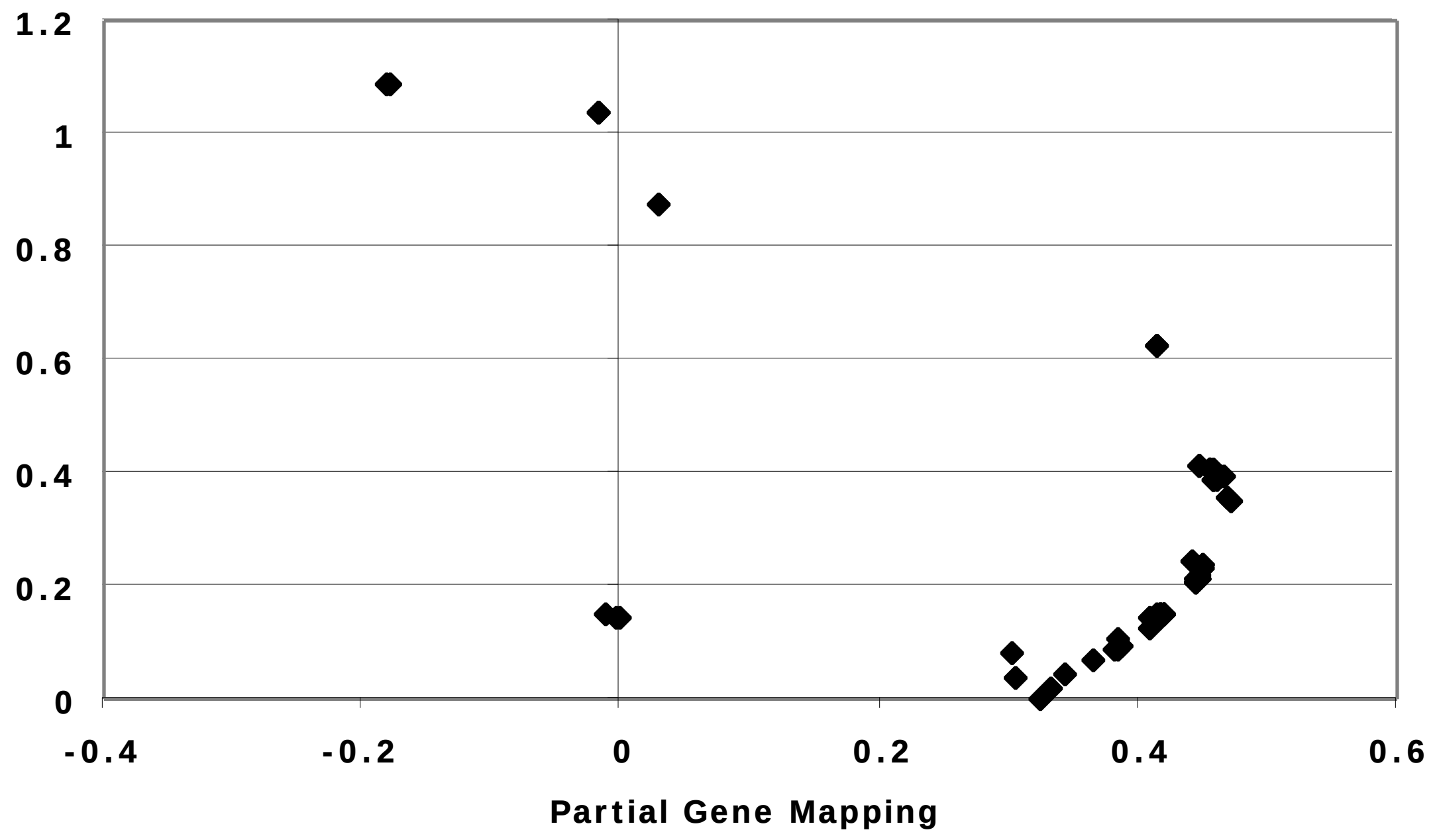

## VI. Computational Efficiency

Although many points are being correlated against each other, the clustering generally limits the correlation computations to much less than $O(N^2)$. The underlying model generation scales linearly with the number of samples and the computation can be performed incrementally, with new data being processed as is it created. Model generation is an almost perfectly parallel process also. ECLM has similar computational advantages to a recently developed technique call fractal clustering [68]. This method has been used to analyze patterns on the Internet [71]. Although parts of the analysis discussed in the paper were done piecemeal and there was human intervention at points, the entire process is algorithmic and all steps from raw data to network diagrams can be fully automated.

To preprocess and run the genetic data took approximately 9 hours on a PowerMac G4 with dual 450 Mhz processors [69]. The stock data took roughly the same amount of time to run.

# References

*Formerly University of California, San Francisco, sandys@itsa.ucsf.edu